# El Triángulo de Platón y El Factor Gnomónico: Una aplicación a los oráculos de Herodoto

Raúl Pérez-Enríquez
Departamento de Física
Universidad de Sonora
rpereze@correo.fisica.uson.mx

A modification to the “gnomonic factor” using the concept of “Plato’s triangle” is presented. With the aid of the “platonic gnomonic factor” (***fgp***) as we called it, we find that the oracles mentioned by Herodotus in his *History,* Dodona in Greece and Ammon in Oasis Siwa, Libya, were placed there because the noon sun’s shadow of a vertical gnomon formed, back in 2500BC, the Plato’s triangle the former, and the “Egyptian sacred triangle” the latter. This means that both concepts were known by Egyptians form Thebes long before they were formalized by the Greeks. The right angled triangle concept is an idealization, as said by D. Magdolen, of an astronomical observation; i.e., it is the shadow cast by a gnomon.

------

Se presenta una modificación al “factor gnomónico” usando el concepto de “triangulo de Platón”. Con la ayuda de lo que llamamos “factor gnomónico platónico” (***fgp***) nosotros encontramos que los oráculos mencionados por Herodoto en su Historia: Dodona en Gracia, y Ammon en el Oasis Siwa, Libia, fueron ubicados ahí porque, hacia el año 2500AC, la sombra proyectada por un gnomon vertical formó el “triángulo de Platón” en el primero y el “triángulo sagrado Egipcio” el último. Esto significa que ambos conceptos eran conocidos por los Egipcios de Tebas, a decir de Herodoto, bastante antes de estos fueran formalizados por los griegos. El concepto de triángulo rectángulo sería la idealización, como dice D. Magdolen, de una observación astronómica; esto es, la proyección de la sombra por un gnomon.

## 1.- Introducción

Las Matemáticas egipcias tienen un comienzo indefinido. Existen una gran variedad de hipótesis acerca de cómo fue que, primero en Mesopotamia, y más tarde, en Egipto, surgieron las nuevas Matemáticas que se expresaron en la ingeniería [1]. La construcción de las grandes pirámides y edificios hace ya casi 5000 años, implica un nivel de conocimiento de la geometría muy superior al que está registrado en los diversos documentos de la antigüedad (papiros y tabletas de arcilla) [2]. Por otro lado, la Astronomía practicada por dichas culturas ha quedado registrada en diversos medios; entre ellos, las pinturas y bajo relieves en los muros de los templos y construcciones [3].

Un hecho que no podemos dejar de considerar es que en estas primeras culturas, el Sol jugaba un papel primordial al grado de ser considerado el Dios principal; tenían dioses del disco solar, de sus rayos, en fin, de sus diversas manifestaciones [4]. En consecuencia, ellas debieron de tener un conocimiento muy profundo de su movimiento en el cielo a los largo del día; también, a lo largo del año. La observación del Sol, seguramente, les permitía lograr la orientación de sus pirámides [5]; y debe estar reflejada en el Papiro de Rhind [2].

De estas consideraciones acerca de la importancia de la observación del Sol, es que sugerimos hace unos años al estudiar el Monumento de Stonehenge, la definición del concepto de factor gnomónico (***fg***) [6]. La utilización de este concepto que podríamos resumir como el cociente entre la diferencia algebraica de las sombras de un gnomon y la altura del gnomon (ver Figura 1), nos permitió interpretar, entonces, la Herradura de Trilitos de ese monumento, como un calendario de tres estaciones. En la figura se ilustra la construcción del ***fg*** para la latitud de Stonehenge.

Figura 1

Ahora, retomamos dicho concepto para definir el *factor gnomónico de platónico* (***fgp***), gracias a la descripción del Triángulo de Platón que nos presentara Jenofonte Moussas recientemente (*pers. comm.*). Descrito, aparentemente por primera vez por Platón en el *Timeo* (53b7-56e7) [7], este triángulo que mostramos en la Figura 2, posee su cateto adyacente igual a dos unidades. Así, un triángulo rectángulo isósceles sobrepuesto sobre otro triángulo configura este objeto matemático.

Figura 2.

Con estos antecedentes, podemos iniciar la presentación de este trabajo. En la primera sección retomamos la *Historia* de Herodoto en lo que respecta a la ubicación de los oráculos de Dodona, en Grecia, y de Ammon, en Libia: su ubicación y su origen. En la segunda parte, pasamos a considerar el concepto de *factor gnomónico de platónico*, fusión del factor gnomónico y la definición del Triángulo de Platón, para establecer la metodología de nuestro análisis. A partir de ahí, en la tercera sección, discutimos sobre el conocimiento matemático de los Triángulos de Platón y Egipcio como conocidos antes de su formalización por los griegos. En la última parte de este trabajo, hacemos algunas reflexiones sobre estos hallazgos y sus implicaciones que tiene la el método del factor gnomónico platónico para entender la ubicación de las ciudades y orientación de los edificios y construcciones de la antigüedad.

## 2.- Oráculo de Dodona según Herodoto

Herodoto, en su *Historia* [8], nos relata la forma en que los oráculos de Dodona y de Ammon fueron ubicados. En el libro II Euterpe (LIV-LVIII) se puede leer que dos sacerdotisas proveniente de Tebas, en Egipto, fueron quienes indicaron a la gente de Grecia en el primer caso, y de Libia, más específicamente del Oasis de Siwa, en el segundo, los lugares exactos en los que deberían erigirse estos centros de adivinación. También, nos describe la forma en la que "esas palomas negras" (*black doves*) fueron llevadas por comerciantes fenicios a esos parajes.

En efecto, "los oráculos hablaban sus profecías a través del crepitar de las hojas del árbol del roble sagrado, del murmullo del agua que brotaba en el manantial cercano, del aleteo de las palomas y de los sonidos emitidos por el viento cuando golpeaba un arreglo de trípodes de bronce o de la caída de dados de bronce sobre un disco de cobre", escribe Herodoto [9].

A partir de estas indicaciones y de las ruinas que se observan en la actualidad, podemos decir que el oráculo de Dodona, erigido para halagar a Zeus, estaba localizado a 22km hacia el sur de la ciudad de Janina, en Grecia. El lugar que ocupaba, entonces, el roble sagrado tendría una Latitud 39°32'47" N y una Longitud 20°47'16" E. En la Figura 3, mostramos una vista del sitio a través de *GoogleEarth*. El sitio seleccionado para los cálculos corresponde al foro del antiguo teatro de Dodona tal como se puede apreciar.

Figura 3

Por otro lado, tenemos las ruinas del oráculo de Ammon, en el Oasis de Siwa, ubicado al este de Libia, casi en la frontera con Egipto. Como ya mencionamos, Herodoto, liga a los dos oráculos a través de un relato que le hicieron saber tres sacerdotes de Ammon originarios de Tebas. Podemos localizar este templo en las siguientes coordenadas: Lat. 29°12'04" N, Lon. 25°32'42" E. En la zona central del Oasis Siwa se pueden identificar dos conjuntos de ruinas: las de la cima del Aghurmi Hill y las ubicadas un poco más al sur que resultan ser más antiguas. Se seleccionó el sitio arriba descrito en virtud de que K.P. Kuhlmann nos comentó que

> "...I understand Aghurmi Hill as place of the ancient Siwan acropolis (which includes the temple of their god Ammon) to have been a "natural" choice. It is my considered opinion that no "philosophical" or in any way "Greek" deliberations influenced the choice of place for the Temple of the Oracle. The latter is dated (by a royal name in Hieroglyphics) to the reign of Pharaoh Amasis (II; 570-526 BC)…" (Kuhlmann *pers. comm.*)

Sugiriendo buscar una correspondencia de fechas con los cálculos.

Figura 4

Veamos más en detalle este origen de los oráculos:

> "El siguiente relato se puede escuchar aún en Egipto concerniente a los oráculos de Dodona, en Grecia, y de Ammon, en Libia... Ellos dicen [los sacerdotes] 'que dos mujeres sagradas fueron sacadas de Tebas por los Fenicios, y que el cuento fue que una de ellas fue vendida en Libia, y la otra en Grecia, estas mujeres fueron las primeras fundadoras de los oráculos en los dos países'"
>
> (Herodoto 440 BC)

¿Cuál es el elemento en común que tienen estos dos lugares? ¿Qué sabían las dos "mujeres egipcias" que fueron vendidas como esclavas?

## 3.- Factor Gnomónico de Platónico

El factor gnomónico (***fg***) es un concepto que desarrollamos a partir de la observación de los grandes trilitos de Stonehenge. Definido como la diferencia algebraica de las sombras del mediodía de un gnomon en las fechas de las posiciones extremas del Sol en el año (los solsticios) dividida por la altura del mismo gnomon, el ***fg*** resulta un método útil para explorar las culturas de nuestros antepasados; creemos que este método estuvo presente de alguna manera en las culturas adoradoras del Sol en el mundo antiguo. Suponemos que ellas por ser conocedoras del movimiento del Sol de sur a norte y de norte a sur a lo largo del año pudieron identificarlo de alguna manera. Culturas como las mesoamericanas parecen haber tenido noción de este método como se desprende de nuestro análisis del calendario Maya [10].

La definición misma de este factor implica la medición de la sombra de un gnomon vertical y, de ello, se desprende la necesaria comprensión del concepto de triángulo rectángulo. Ya D. Magdolen, en su artículo sobre el origen astronómico del triángulo egipcio esboza la idea de que "At that time of the winter solstice, the shadow's length from the gnomon is recorded and compared to other seasonal culminating moments of the sun, but on this day ...[it] must have been one third longer than the proper gnomon's [length]..." [11].

Ahora bien, debido a que en ningún lugar del planeta podemos llegar a observar a partir de las sombras de un gnomon en los solsticios un triángulo como el de Platón, introducimos una modificación al factor gnomónico para definir un nuevo factor:

> Al factor que se obtiene de calcular la diferencia algebraica de sombras de un gnomon en los días del solsticio de invierno (SI) y el día en el que el Sol se encuentra con una elevación de 45° (día gnomon o ***gd***), lo llamamos *Factor Gnomónico de Platónico* (***fgp***), por estar inspirado en el triángulo de la Figura 2.

Así definido, podemos preguntarnos ¿en qué latitud se observó el triángulo con ***fgp***=1 a partir de la observación del Sol de mediodía? O, en otras palabras, ¿en qué sitio el Sol tenía una altitud de 26.57° en el mediodía del solsticio de invierno? La fecha del ***gd*** se obtendría como resultado (ver Figura 5).

Figura 5.

A continuación, describimos cómo la utilización del método del factor gnomónico platónico pudo haber servido para identificar los sitios donde se ubicaron los oráculos mencionados por Herodoto, muchos años antes de la propia definición del triángulo de platón.

## 4. Aplicación del Método: los oráculos de Herodoto

Utilizando un programa de cálculo de posiciones solares como, por ejemplo, *Solar System Live* [12], encontramos que para el solsticio de invierno del año AD 2010, el Sol tenía una elevación sobre el horizonte de 26.57° cuando nos ubicamos en una latitud de 40.02° N; y, para el 6 de octubre, éste alcanzó una altitud de 44.84° (ver Tabla 1). Con ayuda del *Google Earth* [13], pudimos localizar a la ciudad de Janina, en Grecia, como el sitio más cercano. Pero, aparentemente, la antigüedad de esta ciudad no se remonta a la época de Herodoto, ni mucho menos. Sin embargo, al investigar sobre esta ciudad, descubrimos que a solo 22km, se localizan las ruinas de lo que fue el primer oráculo griego: el oráculo de Dodona.

Tabla 1.

En esa misma tabla, mostramos las altitudes solares calculadas para las coordenadas (latitud, longitud) del Oráculo de Dodona descrito en la sección 2. Debemos destacar que dichas elevaciones se obtienen para una fecha de hace 4000 años; esto es, para el día 6 de enero de 2000 BC; esto es, la observación del Sol en Dodona puede considerarse como la realización objetiva del concepto matemático Triángulo de Platón y el ***fgp***=1.

La aplicación exitosa del método del factor gnomónico platónico y la coincidencia de que Herodoto mencionara un segundo oráculo importante de la antigüedad, nos llevó de inmediato a estudiar el triángulo observado en el Oasis de Siwa, también, perteneciente a una fecha remota (2000 BC). En la Tabla 2, mostramos que la sombra del Sol de mediodía en el oráculo de Ammon es de 1.333 unidades suponiendo un gnomon de una unidad de altura. Este resultado expresado en términos de gnomónicos es que ***fgp***=1/3.

Tabla 2.

En la Figura 6, podemos observar que este triángulo formado por el gnomon y sus sombras es nada más ni nada menos que el denominado “triángulo sagrado” egipcio de lados 3:4:5, que es ampliamente conocido como Triángulo Pitagórico pues es el triángulo más sencillo que cumple con la condición del teorema de Pitágoras.

Figura 6.

En consecuencia, consideramos que los dos oráculos a referidos por Herodoto en *Euterpe* tienen al menos tres condiciones básicas en común: i) están ubicados de acuerdo con instrucciones dadas a los pobladores por sacerdotisas provenientes de Tabas, en Egipto; ii) la observación del sol en ellos reproduce triángulos similares al Triángulo de Platón; y, iii) los ***fgp*** de ambos oráculos son fracciones unitarias. La primera observación esta explícitamente dicha por Herodoto. Respecto de la segunda y la tercera, podemos decir que son resultado de lo expuesto en este trabajo. En particular, en la tercera se refiere a que 1 es el recíproco de 1 y 1/3 es el recíproco de 3; por tanto, la condición se cumple.

## 5. Conclusiones

Se ha presentado el método del *factor gnomónico platónico* como método de estudio de ubicación de ciudades y como posible método aplicado por las culturas de la antigüedad que conocían los movimientos del Sol en el cielo a lo largo del año. Asimismo, se muestra que los conceptos matemáticos ahora conocidos como "triángulo de Platón" y "triángulo sagrado egipcio", pueden ser vistos como conceptos derivados de la observación del Sol utilizando un gnomon vertical. Apoyándonos en el *factor gnomónico de platónico*, ***fgp***, podemos decir que dichos triángulos tienen un ***fgp*** de 1 y 1/3, respectivamente. La ubicación de dos importantes oráculos de la antigüedad, mencionados por Herodoto en su *Historia* (el de Dodona y el de Ammon en Libia), en lugares en los que hace alrededor de 4000 años se observaron estos triángulos, implica que dichos conceptos matemáticos ya eran manejados en Egipto; y, que probablemente, las sacerdotisas que salieron de Tebas, conocían o iban en busca, de aquellos parajes en los que dichos triángulos se debían formar. De alguna manera, debemos releer a Herodoto y otros clásicos pues creemos que estos resultados deben aparecer reflejados en los conocimientos matemáticos de los egipcios que están registrados, por ejemplo, en el Papiro de Rhind o en las ciudades construidas a lo largo de los valles del Nilo.

## 6. Agradecimientos

Queremos agradecer, en primer término, a la Universidad de Sonora por el apoyo brindado para la realización de este trabajo; en segundo término, a Xenophon Moussas por habernos acercado al conocimiento del Triángulo de Platón.

**7. Referencias**

## Caption Figures

**Figura 1. Definición del Factor Gnomónico. Dos triángulos rectángulos definidos por el gnomon, su sombra y el rayo de luz proveniente del Sol; ambas sombras corresponden a las posiciones extremas del Sol en el año (los solsticios en Stohehnege).**

**Figura 2. El Triángulo de Platón. Conformado por dos triángulos rectángulos: uno isósceles; el otro con catetos 1:2.**

**Figura 3. Fotografía aérea del oráculo de Dodona (Google Earth). Se toma el centro del foro como punto de ubicación de la observación del triángulo de Platón en el año 2000 BC.**

**Figura 4. Fotografía aérea de las ruinas del antiguo templo de Ammon en el Oasis Siwa, en Libia (Google Earth). Al año 2000 BC, el gnomon y sus sombras formaron el triángulo sagrado egipcio (lados 3:4:5). La fecha en que el Sol alcanzó los 45° fue el 20 de Noviembre.**

**Figura 5. Altitudes del Sol para el Triángulo de Platón, donde *fgp*=1- Este triángulo se formó en el año 2000 BC con los rayos solares del solsticio de invierno y el día en el que el Sol alcanzó 45° como máxima altura al mediodía (octubre 23)**

**Figura 6. Triángulo observado en el templo de Ammon, en el Oasis Siwa, hace 4000 años. Es fácil comprobar que se trata del "triángulo sagrado" egipcio.**

## Tables

| Table 1. Sun's Altitude to form the Plato's Triángle[a]. | | | | | | |
|---|---|---|---|---|---|---|
| Date | Ángle | Year AC 2010 (Lat. 40.02° N) | | Year 2000 BC (Lat. 39.53° N) | | Difference 2010 / -2000 |
| Winter Solstice (WS) | 26.57 | 26.57 | 22-Dec | 26.58 | 06-Jan | 0.00 / 0.01 |
| Gnomon Day (GD) | 45.00 | 44.84 | 06-Oct | 44.80 | 23-Oct | 0.12 / 0.20 |
| Summer Solstice (SV) | | | | 74.40 | 24-Jun | |
| [a], Ángles in degrees. | | | | | | |

| Table 2. Herodoto's Ancient Oracles Analysis | | | | | | |
|---|---|---|---|---|---|---|
| Location | Lat. (°) | Lon. (°) | Date (BC) | Alt.[a] | Shadow[b] | ***fgp***[c] |
| Siwa Oasis, Lybia | 29.20 N | 25.55 E | 01/06/2001 | 36.877 | 1.33299 | |
| Oracle of Ammon | | | 11/20/2000 | 45.148 | 0.99485 | 1/3 |
| | | | | | | |
| Oracle of Dodona, | 39.55 N | 20.78 E | 01/06/2001 | 26.527 | 2.00333 | |
| Greece | | | 10/23/2000 | 44.864 | 1.00476 | 1 |
| [a], Altitude of Sun for wintwr solstice and gnomon day.<br>[b], using a gnomon of unit length<br>[c], platonic gnomónic factor (ver texto) | | | | | | |

Figuras

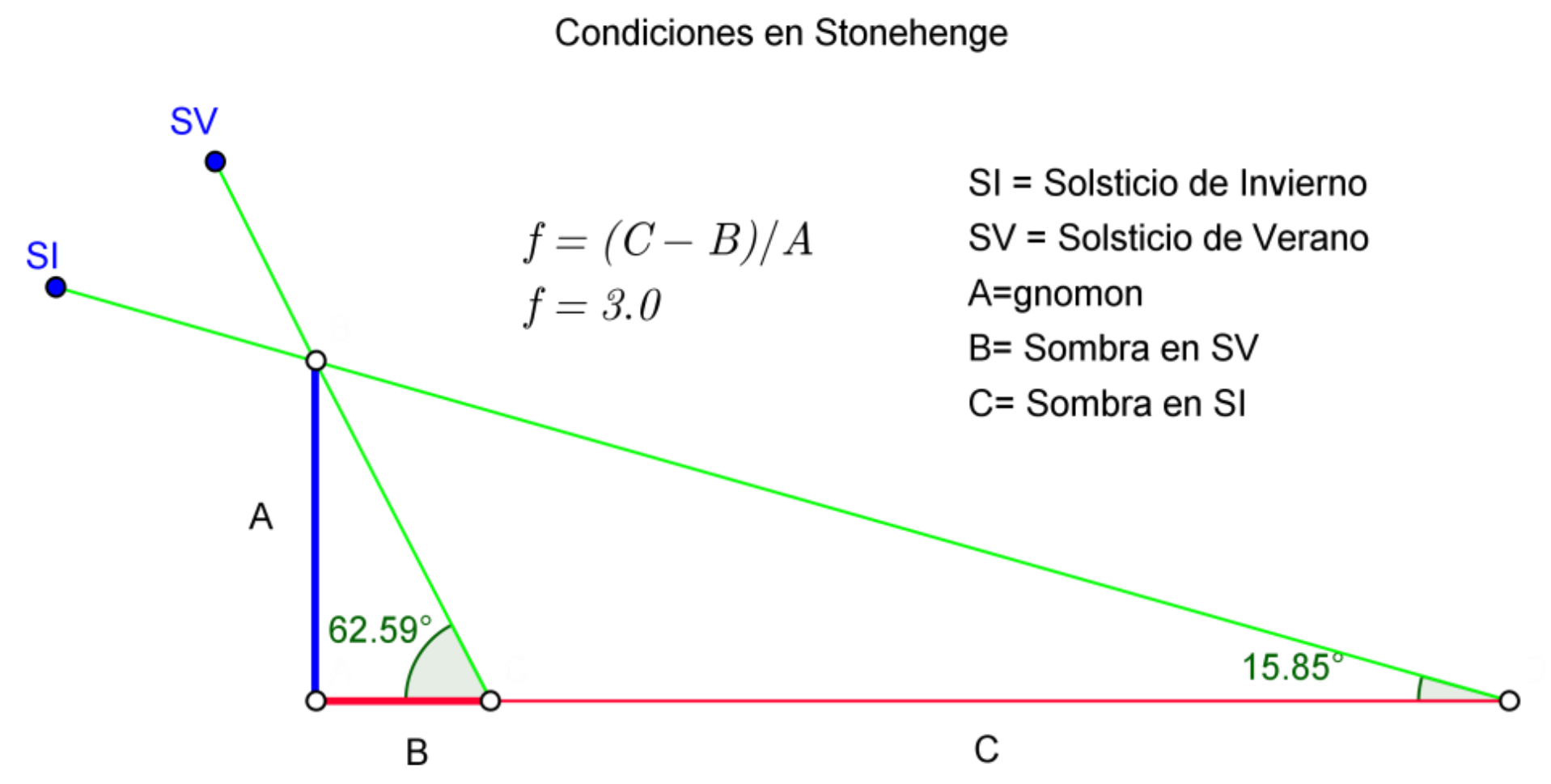

Figura 1.

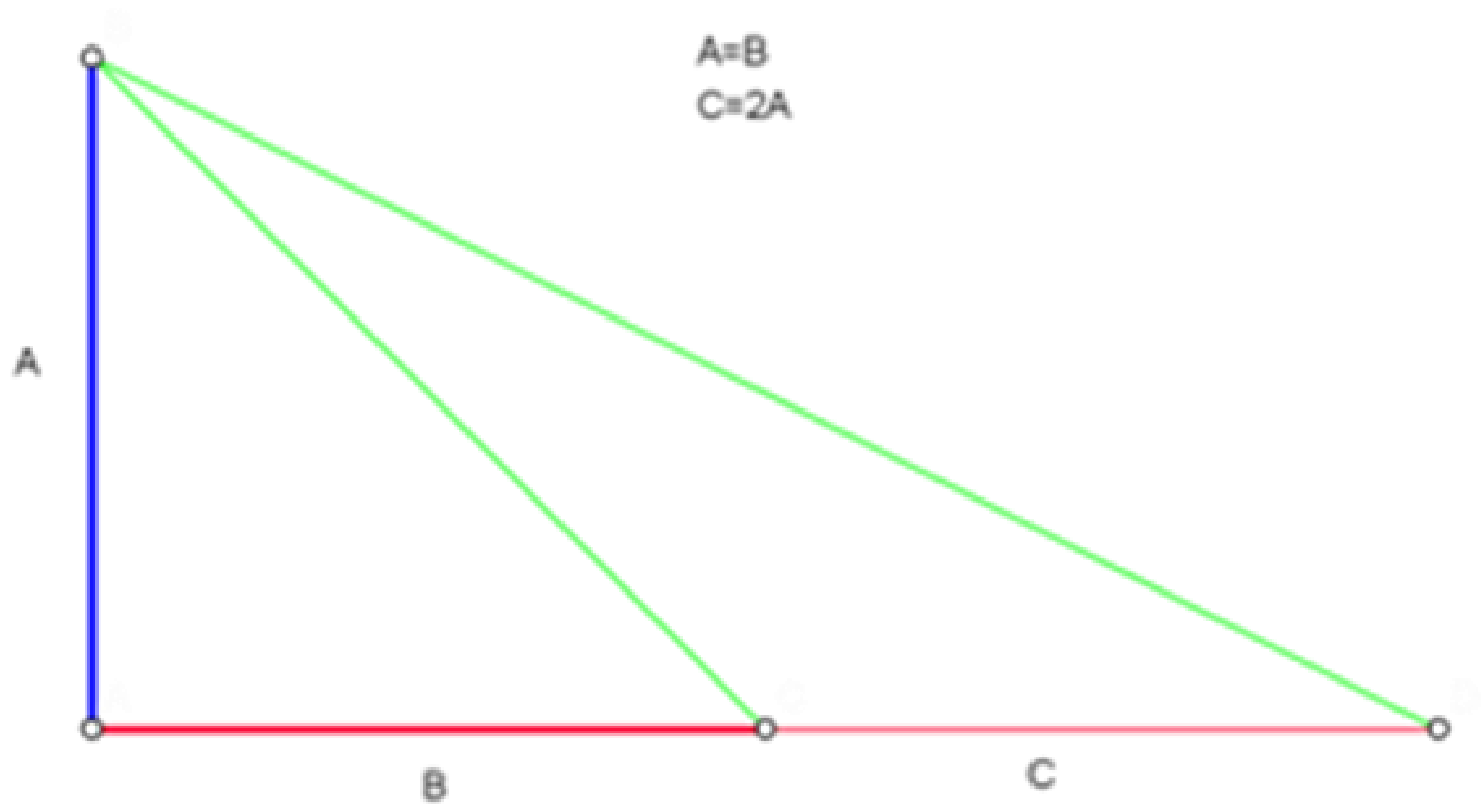

Figura 2

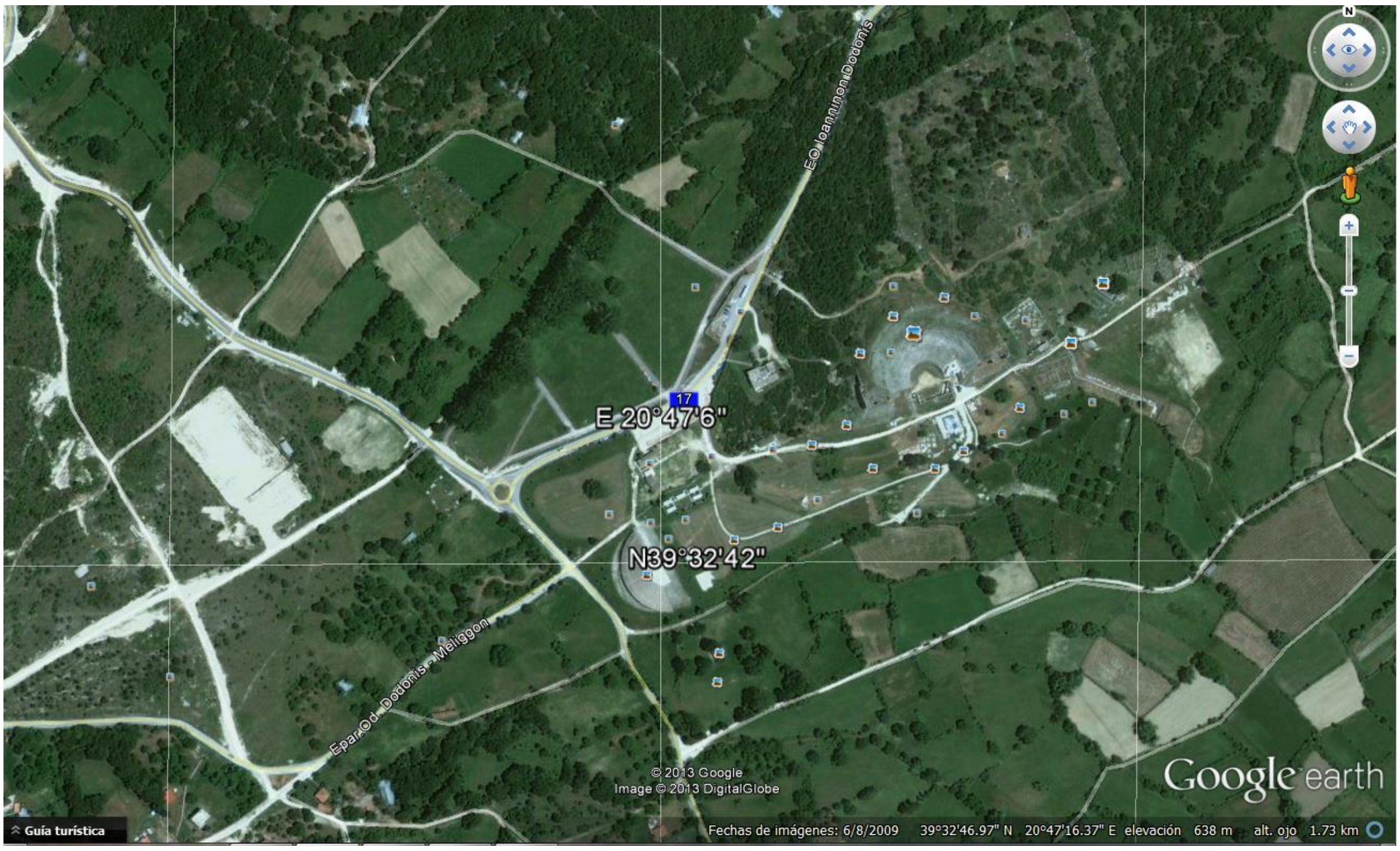

Figura 3

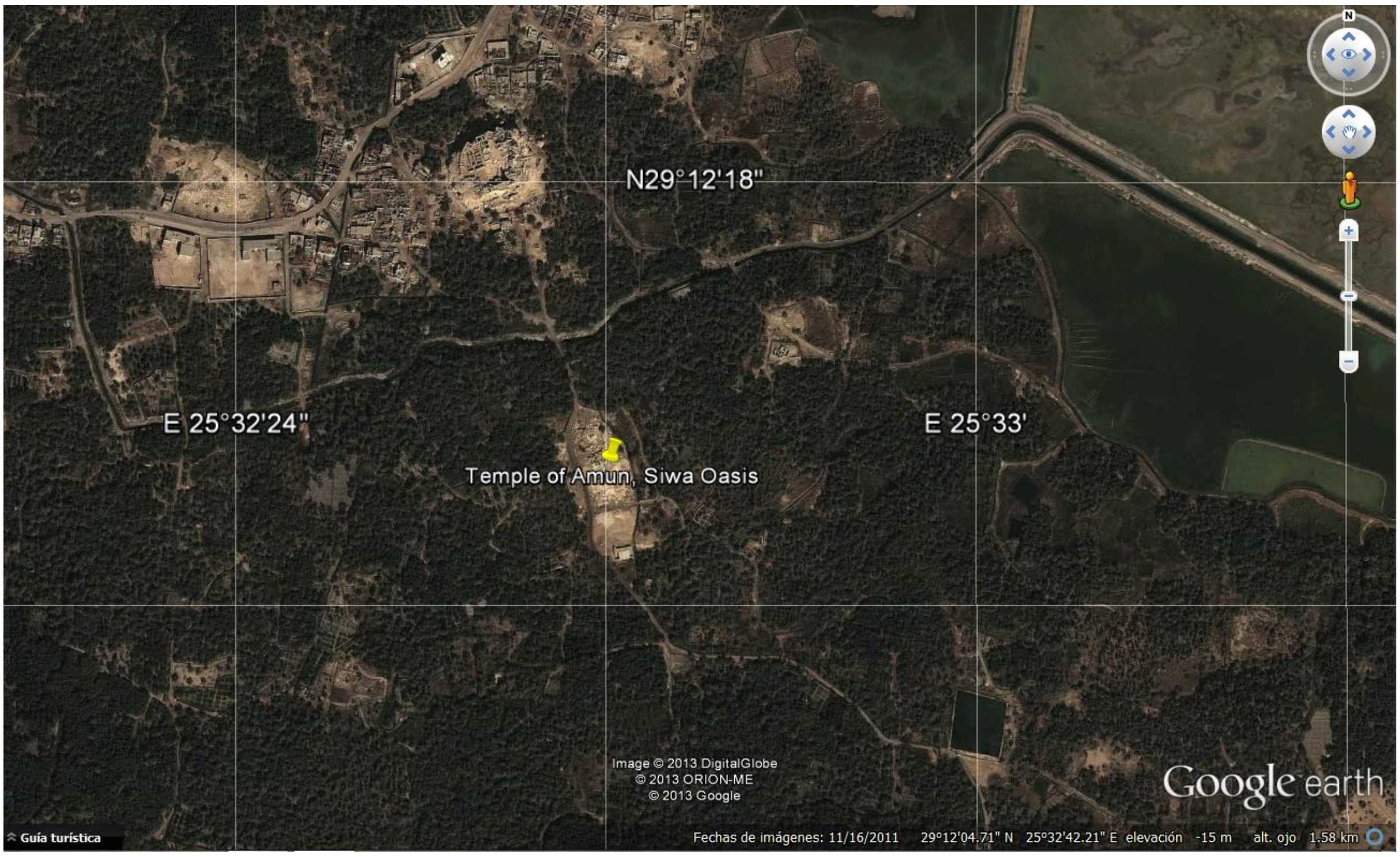

Figura 4

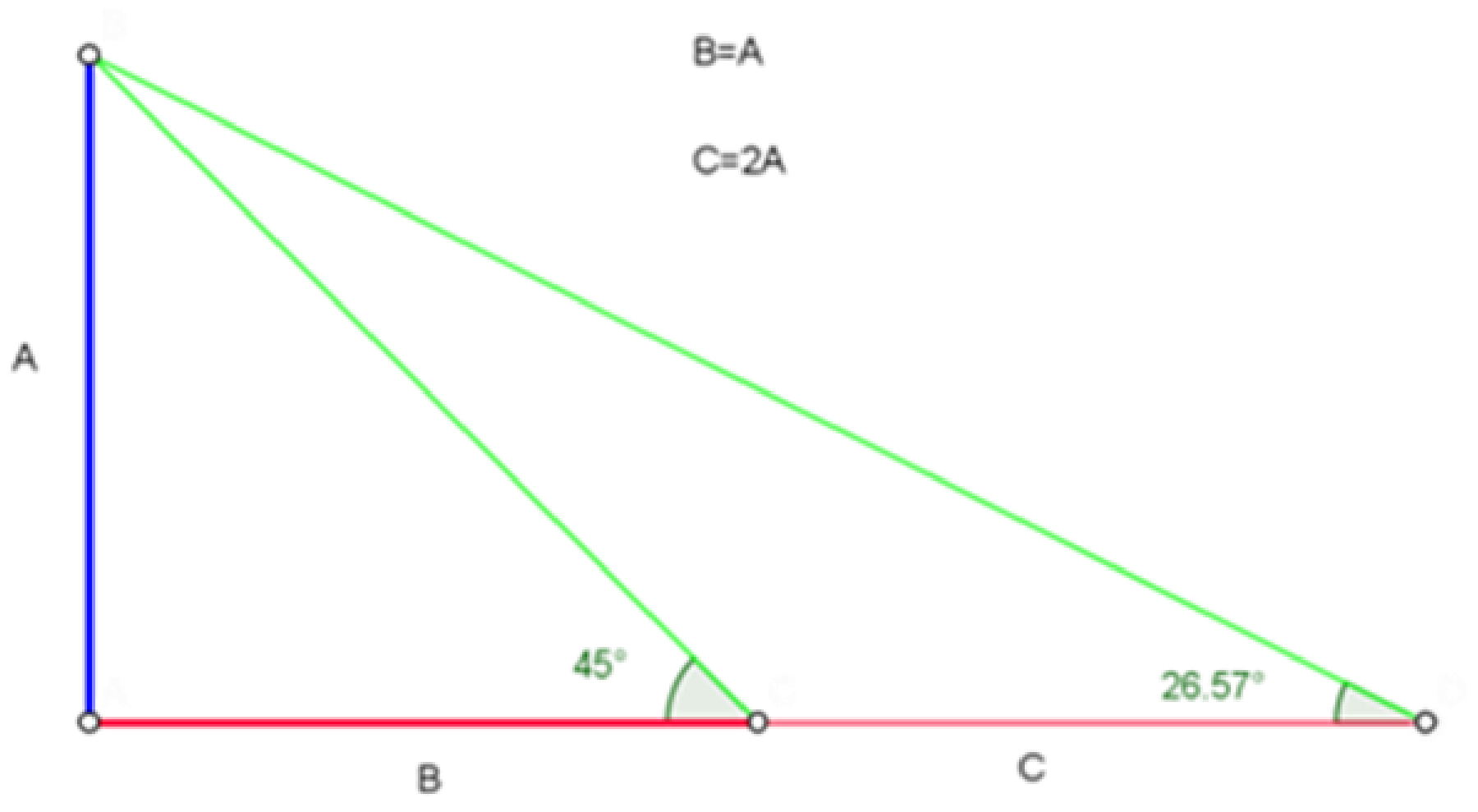

Figura 5

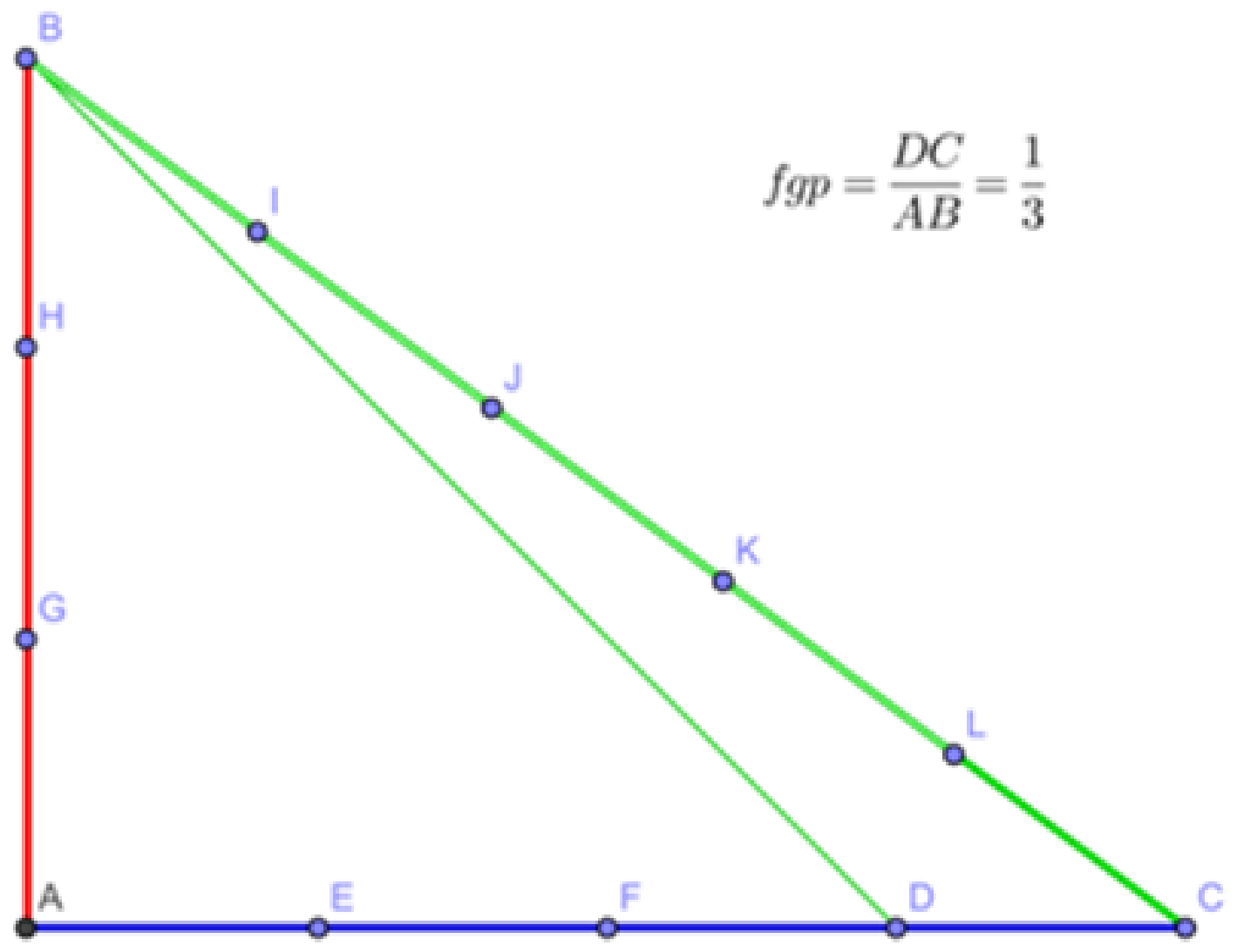

Figura 6.